\documentclass[aps,nofootinbib,superscriptaddress,twocolumn,prl,longbibliography]{revtex4-1}
\usepackage{amsmath,amsthm,amsfonts,amssymb,bm}
\usepackage{graphicx,color}
\usepackage[colorlinks={true}]{hyperref}
\hypersetup{citecolor={blue}, filecolor={blue}, linkcolor={blue}, urlcolor={blue}}
\usepackage[caption=false]{subfig}
\usepackage{epsfig,times}

\def\Tr{\mbox{Tr}}
\begin{document}
\title{Phase-space interference in extensive and non-extensive quantum heat engines}
\author{Ali \"{U}. C. Hardal}
\affiliation{Department of Photonics Engineering, Technical University of 
Denmark, \O rsteds Plads 343, DK-2800 Kgs. Lyngby, Denmark}
\author{Mauro Paternostro}
\affiliation{Centre for Theoretical Atomic, 
Molecular and Optical Physics, School of Mathematics and Physics, Queen's University, 
Belfast BT7 1NN, United Kingdom}
\author{\"{O}zg\"{u}r E. M\"{u}stecapl{\i}o\u{g}lu}
\affiliation{Department of Physics, Ko\c{c} University, \.Istanbul, 34450, Turkey}
\email{omustecap@ku.edu.tr}
\date{\today }
\begin{abstract}
Quantum interference is at the heart of what sets the quantum and classical worlds apart. 
We demonstrate that quantum interference effects involving 
a many-body working medium is responsible for genuinely non-classical features in the performance of a quantum heat engine. The features with which quantum interference manifests itself in the work output of the engine depends strongly on the extensive nature of the working medium. While identifying the class of work substances that optimize the performance of the engine, our results shed light on the optimal size of such media of quantum workers to maximize the work output and efficiency of quantum energy machines. 
\end{abstract}


\maketitle


Quantum interference is responsible for fundamental differences between quantum and classical dynamics~\cite{the_kamland_collaboration_precision_2008,werner_observation_1975,colella_observation_1975,ehrenberg_refractive_1949,aharonov_significance_1959}.
However, as it might be naively believed, the effect of interference is not always to provide advantages and it can also work ``against quantumness".  According to the path integral
formulation of quantum mechanics, constuctive interference of quantum paths
makes them converge to the classical trajectory in the limit of $\hbar\rightarrow 0$~\cite{feynman_quantum_2010}. It is thus necessary to determine proper parameter regimes for a given system such that the occurrence of interference can
be taken as an unambiguous signature of the quantum character of the system.

\par

In this paper we consider a system of $N$ spin-$1/2$ particles interacting through either extensive or non-extensive versions of 
the anistropic Lipkin-Meshkov-Glick (LMG) model~\cite{lipkin_validity_1965}, which is central to many physical problems~\cite{unanyan_decoherence-free_2003,lee_adiabatic_2006,micheli_many-particle_2003,milburn_quantum_1997,kitagawa_squeezed_1993,morrison_dynamical_2008}. 
In the extensive version of such model, the mean energy per spin is finite, while in the non-extensive one it becomes infinite~\cite{campa_statistical_2009}. We use such formulations of the Hamiltonian of our $N$-spin system to describe the working fluid of a quantum heat engine~\cite{quan_quantum_2007,kosloff_quantum_2014}, assessing the effects that  (non-)extensiveness has on the capability of the engine to produce work.

\par

The motivation for such a study stems from a
fundamental question: ``{\it Can we find genuine interference-induced quantum advantages, scaling up with the number of quantum 
workers, in the energetics of quantum many-body systems}"? 
The answer to this question, which is central to emerging fields such as quantum biology~\cite{lambert_quantum_2013} and  quantum thermodynamics~\cite{kosloff_quantum_2013}, can have practical relevance for the optimization of the performance 
of quantum heat engines and bioenergetic systems such as artificial light harvesting complexes~\cite{jesenko_optimal_2012}. 

\par

We show that fundamental interference-related differences in the performance of engines result from the use of extensive and 
non-extensive work media. Moreover, we illustrate the possibility to achieve significant advantages using certain 
parameter regimes of  non-linear interactions, in non-extensive work substances.  
In particular, we find that when the spins in a work medium
interact, this can lead to non-monotonic dependence of the returned work on the number of (working) spins.
The effect stems from quantum interference, which can be geometrically and semi-classically explained in the spin
phase space~\cite{schleich_oscillations_1987,schleich_area_1988,lassig_interference_1993}. 
Specifically, we consider a quantum Otto cycle~\cite{quan_quantum_2007} and find out that interference
is manifested in the work output of the engine ($W$) through oscillations that depend on the parity of the number of of spins in the work medium.
We determine the parameter regimes and approach to discern the contribution to such oscillations that 
can be uniquely associated with phase-space
interference. Moreover, we examine the efficiency and the dependence on $N$ of the marginal productivity ($W/N$), which both help to further compare extensive and non-extensive quantum heat engines (QHEs). The behavior of the marginal productivity of a quantum work
substance can be compared to the ``law of diminishing returns'' in micro-economics~\cite{taggart_economics_2003}.
We identify points of maximum and negative productivity,
which can be used to determine the optimal size of the work substance in both the extensive and non-extensive case.
From a practical point of view, such results can be significant for the identification of a cost-efficient scaling law of the productivity of QHEs,
complementing what has been achieved using engineered quantum heat baths~\cite{hardal_superradiant_2015,turkpence_photonic_2017}. 
From the fundamental perspective, instead, phase-space 
interference can be used to witness the quantum character of a QHE, thus contributing to the current open investigation, in this respect, in the field of quantum 
thermodynamics~\cite{uzdin_equivalence_2015}. 

\par


We consider two different (non-extensive and extensive) descriptions of 
the working substance of a quantum Otto engine by
using an anisotropic LMG-type Hamiltonian for a set of $N$ (pseudo)spin-$1/2$ particles, in the absence of any longitudinal magnetic field. 
The model reads (we choose units such that $\hbar=k_B=1$ throughout the paper, with $k_B$ the Boltzmann constant)
\begin{equation}\label{eq:HssNonExtensive}
\hat{H}_{\text{LMG}}=\gamma_x  \hat{S}_x^2+\gamma_y  \hat{S}_y^2,
\end{equation}
where we have introduced the collective spin operators
\begin{equation}\label{eq:collectiveSpinOps}
\hat{S}_{\alpha }=\frac{1}{2} \sum _{i=1}^N \hat{\sigma }_{\alpha },~~~(\alpha=x,y,z).
\end{equation}
Eq.~(\ref{eq:HssNonExtensive}) can describe both the extensive and non-extensive version of the Hamiltonian model, the latter being 
retrieved by performing the scaling $\gamma_{x,y}\to\gamma_{x,y}/N$ of the coefficients. 
Simulations of the LMG model have been proposed based on 
a Bose-Einstein condensate trapped in a double-well potential~\cite{milburn_quantum_1997}, circuit quantum 
electrodynamics~\cite{larson2010circuit}, single-molecule magnets~\cite{campos2011single}, and toroidal condensates~\cite{opatrny2015spin}. 
The model has rotational and spin-flip symmetries. For
simplicity, we will limit our study to the subspace of symmetric Dicke states corresponding to the maximum total 
spin sector with $S=N/2$. In such a manifold, we look for the differences between extensive and 
non-extensive model in the behavior of the mean energies against the particle number. 
While, in general, non-extensive models give infinite mean energy
per spin in the thermodynamical limit (i.e. for $N\to\infty$), extensive ones lead to finite mean energy per spin in such a limit. 
According to the so called Kac rescaling~\cite{kac_van_1963}, extensivity is
recovered at the cost of losing the additivity of the energy, so that the sum of the energies of the 
subsystems can be different from
the total mean energy of the system~\cite{campa_statistical_2009}. 
While in order to study the phenomenology of non-extensive systems, special forms of 
entropy have been proposed~\cite{tsallis_possible_1988}, here we rely on a thermodynamic approach based on the 
performance of a QHE.

\par

We restrict the analysis to the antiferromagnetic model and thus assume $\gamma_{x,y}>0$. The choice of no external magnetic 
field makes the model in Eq.~(\ref{eq:HssNonExtensive}) non critical. Phase transitions can lead to non-extensive behavior in 
heat engines irrespective of their classical or quantum character, which can be used to operate an engine at Carnot efficiency 
with finite power~\cite{campisi_power_2016}. Our purpose here is
to use quantum interference to reveal quantum signatures of the working substance with extensive and non-extensive descriptions. 
Recently, quantum and classical interference have been studied in the ferromagnetic LMG model from the Landau-Zener
perspective, finding significant effects arising from classical rather than quantum interference, the latter being swept out by 
quantum fluctuations~\cite{larson_interaction-induced_2014}. 
In the following, we describe the engine cycle and explicitly show how interference contributes
to its work output.

\par

\begin{figure}[tbp]
\centering
\includegraphics[width=8 cm]{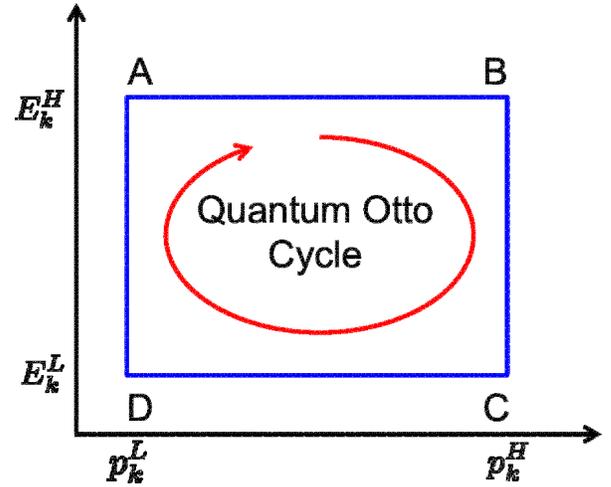}
\caption{Quantum Otto engine cycle in the eigenenergy ($E_k$) and the occupation probability $p_k$ space
of an arbitrary energy level $k$ of the spin system. In the quantum adiabatic stages, $D\rightarrow A$ and $B\rightarrow C$, 
the system is disconnected from the
heat baths and model parameters $\gamma_x,\gamma_h$ changes between $\gamma_x^L,\gamma_y^L$
and $\gamma_x^H,\gamma_y^H$. In the isochoric stages, $A\rightarrow B$ and $C\rightarrow D$, the system is coupled to 
a hot bath at
temperature $T_H$ and a cold bath at temperature $T_L$, respectively, so that $p_k$ changes between $p_k^L$ and $p_k^H$.}
\label{fig:cycle}
\end{figure}


The quantum Otto cycle can be described by an energy-probability $E-p$ diagram depicted
in Fig.~\ref{fig:cycle}, where $p_k$ represents the occupation probability of an arbitrary
energy level $k$ of the model Hamiltonian. It is a four stroke irreversible engine that consists
of two quantum adiabatic and two quantum isochoric stages~\cite{quan_quantum_2007}.
In the adiabatic stages, $D\rightarrow A$ and $B\rightarrow C$, 
the model Hamiltonian changes by varying the parameters $\gamma_{x,y}$ to $\gamma_{x,y}^H$ 
and $\gamma_{x,y}^L$, respectively.
In the isochoric stages, the Hamiltonian is fixed
but the system is in contact with a hot bath at temperature $T_H$ and a cold bath at temperature $T_L$,
so that each energy level population can change to new values of
$p_k^H$ and $p_k^L$, respectively. The internal energies at the end of each stage can be calculated
as $U_\alpha=\Tr(\rho_\alpha \hat H_\alpha)$, where $\rho_\alpha$ and $\hat H_\alpha$ are the density matrix of the working substance
and the Hamiltonian at the point $\alpha=A,B,C$ and $D$, respectively. The relevant thermodynamic quantities of the engine are given by~\cite{quan_quantum_2007} 
\begin{eqnarray}\label{eq:workEffExact}
Q_{\text{in}}&=&\sum_k E_{k}^{H}(P_k^H-P_k^L),~~Q_{\text{out}}=\sum_k E_{k}^{L}(P_k^L-P_k^H),\nonumber\\
W&=&Q_{\text{in}}+Q_{\text{out}},~~\eta={W}/{Q_{\text{in}}},
\end{eqnarray}
where the heat in-take and out-take heat are denoted by $Q_{\text{in}}$ and $Q_{\text{out}}$, respectively (we use the convention according to which $Q_{\text{out}}<0$). In a quantum adiabatic process, eigenstates change
while the populations remain the same. The multilevel
system does not need to be assigned neither a real nor an effective temperature at points $A$ and $C$ as it is allowed be a 
non-thermal state at the end of the quantum adiabatic transformation. 
In our analysis we do not consider finite-time cycles and assume, by the introduction of 
infinitesimal perturbations, sufficiently slow quantum adiabatic transformations. 
In addition we further assume that the target thermal states at points $A$ and $C$ can be achieved at any interaction
strengths $\gamma_x,\gamma_y$ by contact with corresponding reservoirs. The study of quantum interference in the 
output of a more realistic engine is left to future explorations. In order to discern the contribution of quantum interference to the 
work output of the cycle, 
we use perturbation theory for the calculations
of the internal energies $U_x$~\cite{niklasson_density_2004}.

\par


We consider $\gamma_x^H>\gamma_y^H\gg\gamma_x^L>\gamma_y^L>0$. The choice of small values for
parameters $\gamma_x,\gamma_y$ allows us to calculate the 
internal energies 
perturbatively to reveal
the interference physics behind the problem. The ordering of the energies, in particular the choice
$\gamma_x^L>\gamma_y^L$ leads to a constructive interference in the work output, as we 
discuss later on. 
We can thus write~\cite{niklasson_density_2004} 
$U_\alpha\approx \Tr(\rho_\alpha^{(0)}H_\alpha^{(0)})+\Tr(\rho_\alpha^{(0)}H_\alpha^{(1)})$
with 
$\rho_\alpha^{(0)}=\sum_n P_{n}^\alpha|n\rangle\langle n|$,
$H_\alpha^{(0)}=\sum_n \gamma_x^\alpha n^2|n\rangle\langle n|$, and 
$H_\alpha^{(1)}=\sum_m \gamma_y^\alpha m^2|m\rangle\langle m|$.
Here 
$|n\rangle$ and $|m\rangle$ are the angular momentum states for the quantization axis $x$ and $y$ respectively, such that $\hat S_x|n\rangle=n| n\rangle$, and $\quad\hat S_y|m\rangle=m| m\rangle$
with $n,m=-S,-S+1,...,S-1,S$. With this notation
\begin{eqnarray}
U_\alpha\approx\sum_n P_{n}^\alpha \gamma_x^\alpha n^2
+\sum_{n,m} P_{n}^\alpha  \gamma_y^\alpha m^2 P(n,m)
\label{eq:IntEn}
\end{eqnarray}
with the transition probabilities $P(n,m)
=|\langle n|m\rangle|^2$  
and the occupation probabilities $P_{n}^\alpha$ ($\alpha=B,D$)
that are obtained from the canonical distribution
\begin{equation}
\label{eq:PnB}
P_{n}^B=\frac{e^{-n^2\gamma_x^H/T_H}}{\sum_n e^{-n^2\gamma_x^H/T_H}},P_{n}^D=\frac{e^{-n^2\gamma_x^L/T_L}}{\sum_m e^{-n^2\gamma_x^L/T_L}}.
\end{equation}
The output work can thus be written as
$W=W_x+W_{xy}$,
where
\begin{eqnarray}
W_x&=&\sum_n(P_n^B-P_n^D)(\gamma_x^H-\gamma_x^L)n^2, 
\end{eqnarray}
and
\begin{eqnarray}\label{eq:workxy}
W_{xy}=\sum_{n,m}(P_n^B-P_n^D)(\gamma_y^H-\gamma_y^L)m^2P(n,m).
\end{eqnarray}
The term $W_{xy}$ shows the quantum interference contribution to the work output. It involves the transition probabilities between spin states belonging to different quantization axes, which interfere mutually with weights proportional to the difference in populations, and signs determined by $\gamma_y^H-\gamma_y^L$. Such interference can thus be either constructive or destructive, therefore influencing the value of the output work $W$. By taking a low-temperature regime, we can safely assume that $n$ is limited to few small values (only the low-lying states are populated). 
We find that the transition probabilities $P(n,\pm S)$ grow as $n$ decreases. Accordingly, 
the negative 
population differences associated with the lower levels are 
weighted more, so that a 
negative value of the difference $\gamma_y^H-\gamma_y^L$ allows for interference that enhances the work output. The
effect can be explained geometrically using semi-classical interferences in the spin phase space, as described in the Supplemental Material~\cite{supplement}.

\par

\begin{figure}[t!]
{(a)}\hskip3cm{(b)}\\
\includegraphics[width=4 cm]{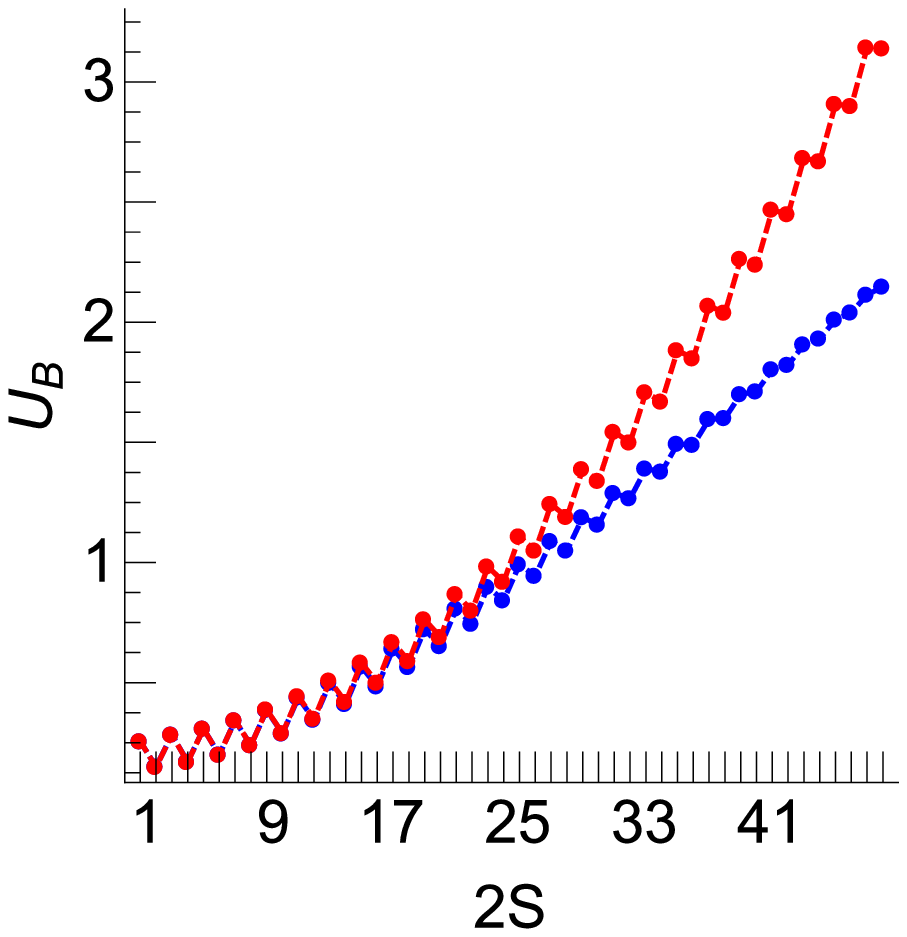}\includegraphics[width=4 cm]{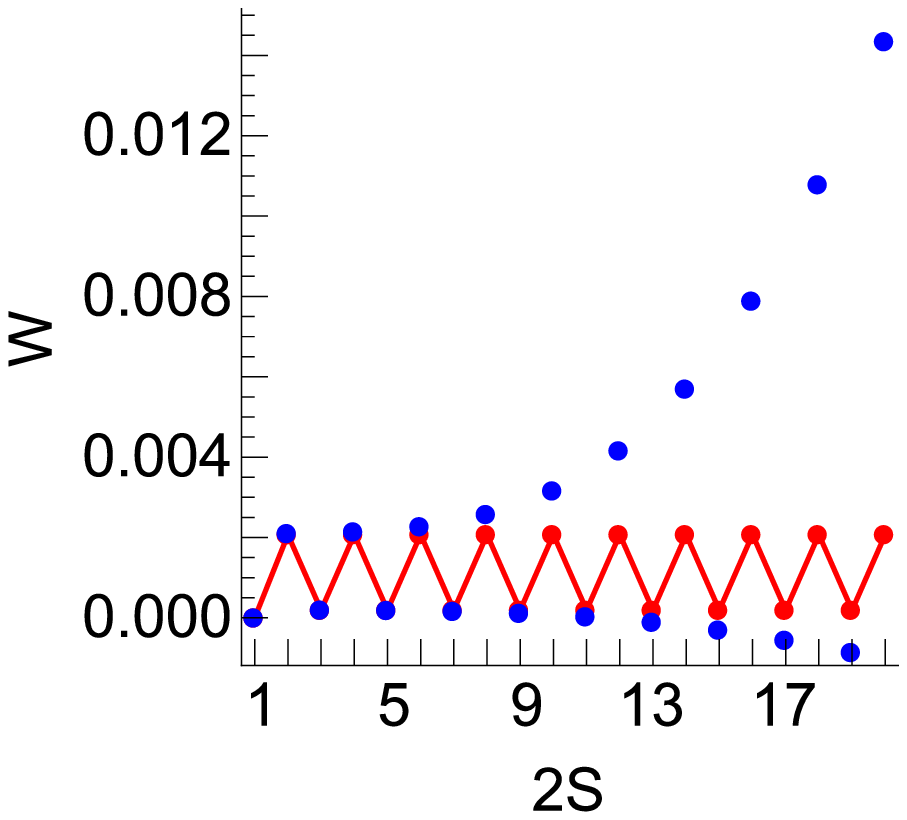}
\caption{Dependence of (a) the internal energy $U_B$ at point B and (b) work output of a quantum Otto 
cycle in Fig.~\ref{fig:cycle} 
with a working substance of interacting spins on the number of spins
$N=2S$ with total spin $S$. The interaction is described by LMG model. 
The upper red curve in (a) and connected red markers (b) are for the perturbative calculation,
while the lower blue curve in (a) and the blue marking points in (b) are for the exact diagonalization method.
We use
dimensionless parameters scaled by $\gamma_x^L$ and take
$\gamma_x^H=1.01$, $\gamma_x^L=1$,
$\gamma_y^H=0.01$ and $\gamma_y^L=0.02$. The temperatures of the heat baths are 
$T_H=0.4$ and $T_L=0.1$. }
\label{fig:exactVSpert}
\end{figure}

For our numerical simulations, we take $\gamma_x^H=1.01$, $\gamma_y^H=0.01$ and $\gamma_y^L=0.02$ (in units of $\gamma_x^L$), which ensure constructive interference.  
We also work in the low temperature limit $T_H\ll\gamma_x^H$ and $T_L\ll \gamma_x^L$ to make any quantum feature more preponderant.
We thus take $T_H=0.4$ and $T_L=0.1$ (in units of $\gamma_x^L$). A word of caution should be spent on the range of validity of the perturbative approach. It works over a 
finite range of total spin $S$, depending on the parameter set used. As far as the the internal energies are considered,
the range of $S$ over which the perturbation method can be applied is relatively wide, 
while, perturbative
evaluations of work is limited to relatively smaller range of $S$, as can be seen in 
Fig.~\ref{fig:exactVSpert}. We calculate the work
output of the engine $W=Q_{\text{in}}+Q_{\text{out}}$,
where
$Q_{\text{in}}=U_B-U_A$ and $Q_{\text{out}}=U_D-U_C$ by exact numerical diagonalization of the model and the perturbative approach. 
The changes in $U_B$ against the number of spins $N=2S$ is plotted in Fig.~\ref{fig:exactVSpert} (a). A similar behavior is found for the other
internal energies at points $A,C,$ and $D$, showing oscillations dependent on the parity of $N$. The oscillations are translated to the work output, plotted in Fig.~\ref{fig:exactVSpert} (b). 

\par

\begin{figure*}[tbp]
\centering
 {(a)}\hskip6cm{(b)}\hskip6cm{(c)}\\
 \includegraphics[width=5.5 cm]{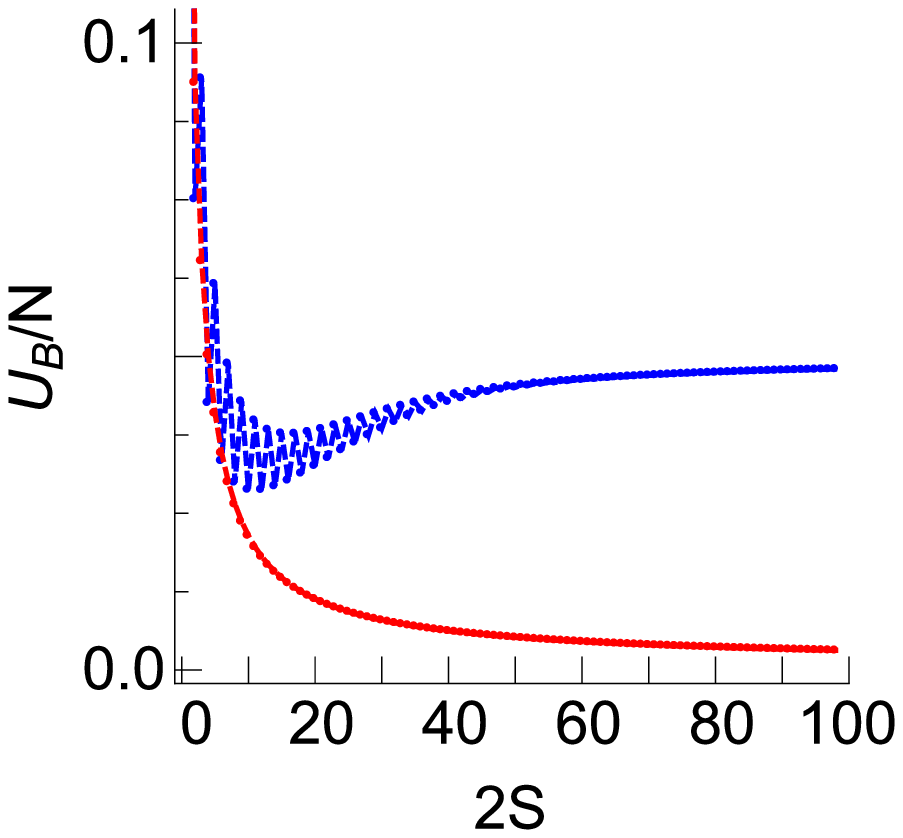}\includegraphics[width=5.5 cm]{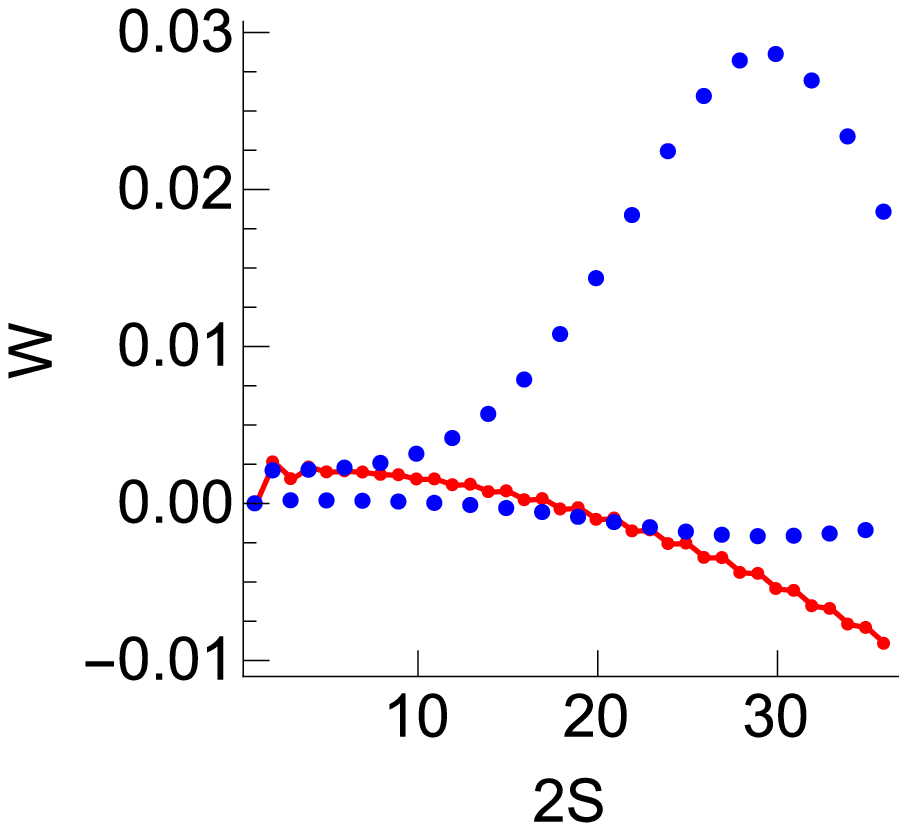}\includegraphics[width=5.5 cm]{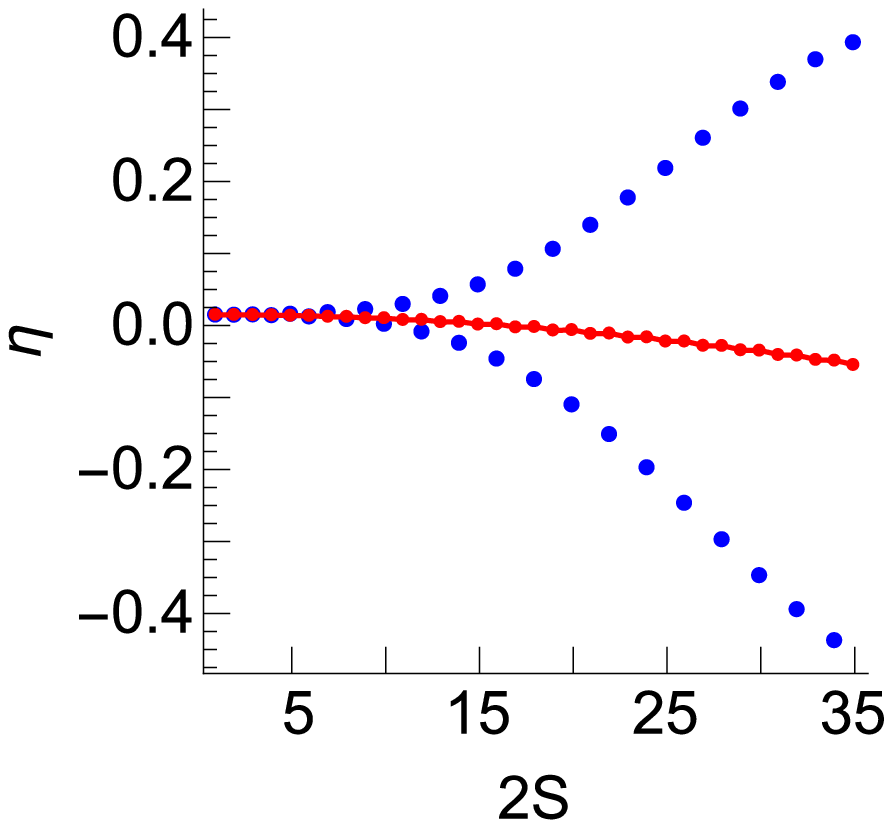}
\caption{Dependence of (a) the internal energy per particle $U_B/N$ at point B, (b) work output $W$, 
and efficiency $\eta$ of a quantum Otto 
cycle in Fig.~\ref{fig:cycle} 
with a working substance of interacting spins on the number of spins
$N=2S$ with total spin $S$. The interaction is described by LMG model. 
The lower red curves in (a) and (b), and the middle red curve in (c) are for the extensive description of the LMG model,
while the upper blue curve in (a) and the blue marking points in (b) and (c) are for the non-extensive description.
We use
dimensionless parameters scaled by $\gamma_x^L$ and take the spin-spin interaction parameters as 
$\gamma_x^H=1.01$, $\gamma_x^L=1$,
$\gamma_y^H=0.01$ and $\gamma_y^L=0.02$. Temperatures of the heat baths are taken to be
$T_H=0.4$ and $T_L=0.1$. }
\label{fig:exactVSnonExt}
\end{figure*}

The growth of $U_B$ with $N$ becomes linear in the thermodynamic limit $U_B\sim N$ or the internal energy per particle
becomes a finite value for large $N$, as shown in Fig.~\ref{fig:exactVSnonExt} (a). 
We remind that our analysis is restricted to the maximum total spin sector of the
Hilbert space. Consideration of the full Hilbert space leads to the expected non-extensive behavior due to the finite energy
left per particle in the thermodynamic limit. The extensive model, applying the Kac's rescaling, 
gives however zero mean energy per particle for large $N$, consistently with the expected 
 thermodynamical limit for an extensive system in the full Hilbert space of $N$ particles. 
 
 \par

The work output according to the non-extensive description of the working
substance shows a super-linear ($\sim N^2$) growth with the even number of spins $N$ up to a critical $N_{\text{max}}\sim 30$ and decrease towards
zero after this point. The behavior of even number of spins is consistent with the micro-economical 
law of diminshing returns~\cite{taggart_economics_2003}. In addition to the ``point of maximum return'', there is a ``point of diminishing return'' at $N_{\text{dim}}\sim 20$
after which each added spin decreases the rate of work production. However, there is a difference in the micro-economy of quantum labor force according to which a non-extensive working system with an odd number of spins does negligible work
or yields even negative returns, while an extensive system can return work. However, even in such latter case, the return is still less than that of the even number of spins [cf.~Fig.~\ref{fig:exactVSnonExt} {\bf (b)}]. 
In the case of an extensive model, $W$ decreases with $N$, exhibiting oscillations dependent on the parity of the latter. Heat-engine operations with significant $W>0$ are only possible for a small number of spins in an extensive model, and the optimum
number of workers is $N=2$.

\par

The efficiency is
evaluated by $\eta=1+Q_{\text{out}}/Q_{\text{in}}$ and plotted in Fig.~\ref{fig:exactVSnonExt} {\bf (c)}. 
Both extensive and non-extensive engines have similar efficiencies for small values of $N$. This can be 
intuitively expected due to the small difference in the
control parameters of the adiabatic stages in our set of parameters. 
Extensive model has negative efficiency after $N\sim 10$ for which the work output is also negative, and the system
cannot operate as a heat engine. Efficiency is weakly showing even-odd oscillations for both models and small $N$. 
The point of maximum return $N\sim 30$ does not coincide with the value of $N\gtrsim 35$ at which the maximum efficiency 
for non-extensive systems is achieved.

\par

The parity-dependent oscillations found in the work and efficiency cannot be immediately associated with the occurrence of quantum
interference.
However, our perturbative analysis led us to determine
a set of parameters for which the oscillations due to spectral differences can be removed from the work output. Accordingly,
we conclude that for sufficiently small number of spins $N<10$ -- over which the perturbation approach is justified --
parity-dependent oscillations are genuinely due to quantum interference. Exact calculations still predict more significant oscillations
for larger number of spins. However, whether these are due to quantum interference or not cannot be deduced from the first-order density
matrix perturbation theory. We leave this as an open problem for future explorations. 

\par


We have considered a system of $N$ spins, interacting through extensive and non-extensive LMG type models, as the working 
system of a quantum Otto engine. Using density matrix perturbation method we have calculated the internal energies, work output, 
and efficiency of the cycle for both interaction models and compared with the exact solutions. We have found oscillations in the work 
output with respect to the number of elements of the working medium, more pronounced for the non-extensive model. 

\par

We have also provided an intuitive comparison with the micro-economical law of diminishing returns. The
work output of a non-extensive engine follows such a law and showcases points of maximum and diminishing returns. 
The parity-dependent oscillations in the output work, however, mark a major difference between quantum and classical working 
media from micro-economical perspective.

\par

Our perturbative approach helped us identify the two sources of these oscillations, which are provided by the differences in the Hamiltonian 
spectrum and the quantum interference between the spin states of different quantization axes contributing to the work output. 
As the latter contribution can be easily isolated, the fine tuning of the perturbation parameter can be used to 
control the (constructive or destructive) character of the interference. 

\par

Quantum interference contribution to the work output can be explained by the interplay of thermodynamics, 
interference in the spin phase space,
and nonlinearity of the spin-spin interaction. 
This suggest that the parity-dependent oscillations in the interference term of the work output associated with the number 
of quantum workers is a genuine quantum thermodynamical effect.

\par

Our results elevate quantum heat engines to the role of fruitful platforms for the fundamental study of quantum interference.
Moreover, the identification of points of maximum and diminishing returns, as quantum analog of the classical law of diminishing returns, 
can be used to optimize the preparation of extensive and non-extensive working substances of quantum heat engines.

\par

%
A.~\"U.~C.~H.~acknowledges support from the Villum Foundation through a postdoctoral block stipend. 
\"O.~E.~M.~and A.~\"U.~C.~H.~acknowledge 
the support by the University Research Agreement between Lockheed Martin Chief Scientist's Office and Ko\c{c} University. M.~P.~acknowledges 
support from the DfE-SFI Investigator Programme (grant 15/IA/2864). \"O.~E.~M. and M.~P.~are grateful to the Royal Society for 
support through the Newton Mobility Grant scheme (grant NI160057). 
%
%
%
%
%

\appendix
\section{Supplemental Material:\\Phase-space interference in extensive and non-extensive quantum heat engines}
\section{Phase-space interference }

\begin{figure}[hb!]
\centering
\includegraphics[width=8 cm]{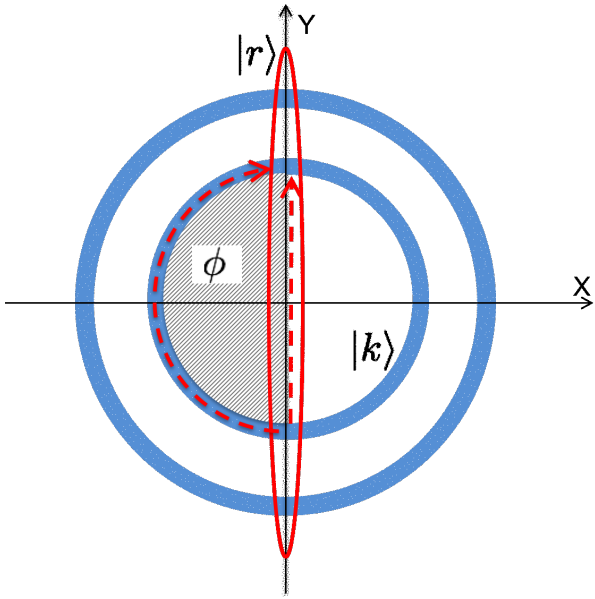}
\caption{Interference of a squeezed vacuum state $|r\rangle$ with Fock number states $|k\rangle$ in the  
oscillator
phase space $X,Y$, where $X$ and $Y$ denote the displacement and momentum, respectively. 
The ellipse and the concentric circular bands represent the squeezed state and the Fock states, respectively.
The ellipse cuts two overlap regions, with area $A=A(k,r)$ 
on the annulus associated with $|k\rangle$. The dashed lines indicate two possible paths connecting
the overlap regions. The shaded area between the dashed lines is denoted by $\phi=k\pi$. The interference can be envisioned 
as a Young interferometer
where the overlap regions correspond to the two slits, or a path interference of the trajectories between the overlap regions,
or superposition of two waves originated from the overlap regions and propagating in opposite directions. 
Transition probability between
the states is determined by $P(k)= 4A\cos^2(\phi/2)$.}
\label{fig:oscillatorPhaseSpcInterf}
\end{figure}

Fock states, $|k\rangle$, where $k$ is an integer, can be depicted
as concentric annuli in the phase space as shown in Fig.~\ref{fig:oscillatorPhaseSpcInterf}. Each band has a finite area
fixed by the semi-classical Bohr-Sommerfeld quantization condition of the action. 
Squeezed vacuum state $|r\rangle$, obtained by 
the squeezing operator, with squeezing parameter $r$, transforming the vacuum Fock state $|0\rangle$, is 
represented by an ellipse. The probability distribution $P(k)=|\langle k | r\rangle|^2$ can be visualized as the 
overlap of the $k-$th circular annulus and the 
ellipse~\cite{schleich_area_1988}. The overlap consists of two regions with finite areas, 
which can be imagined as two slits of a Young interferometer. As the photon number $k$ increases the overlapping regions
get smaller and the distance between them increases so that oscillatory decrease of $P(k)$ with $k$ conforms to the 
double-slit
interferometer behavior. Other interpretations of the interference oscillations can also be considered.
First, two signals from one overlap region following two different paths, propagate to the other overlap region. One path is through the
circular annulus and the other one is  through the ellipse. Signals interfere due to the phase delay between them. Second,
superposition of two waves, emitted from the overlap regions in opposite directions, lead to interference~\cite{schleich_oscillations_1987}.
In these interpretations,
as the slits gets narrower and the distance between them increases with $k$, we expect an oscillatory decrease
of $P(k)$ with $k$, conforming to the analytical calculation showing oscillations in $P(k)$ with respect to
even and odd number of photons~\cite{schleich_oscillations_1987}. The phase delay is
determined by the area between the paths~\cite{schleich_area_1988}. The paths can be approximated
by a line and an arc, respectively, for sufficiently narrow ellipse (large $r$). The radius of the circle is $\sim\sqrt{2k}$, which 
gives the area between the paths to be $\pi k$. The 
interference factor due to this phase difference is given by $\exp{(i\pi k)}=(-1)^k$ so that the probability becomes
$P(k)\sim 1+(-1)^k$.

A phase-space method allows us to use a semi-classical picture to geometrically interpret 
and visualize quantum interferences. A critical signature of interference is the existence of an oscillatory pattern with respect
to some geometrical parameters of the interferometer. In a two slit Young 
interferometer for example, oscillatory behavior emerges with respect to the spatial
separation between the slits, translated into the path difference between different
arms of the interferometer. In quantum optics, oscillations in the
probability distribution of the squeezed vacuum state with respect
to the photon number have been explained using a semi-classical interference picture
in the oscillator phase space~\cite{schleich_oscillations_1987}. 

\begin{figure}[tbp]
\centering
\includegraphics[width=8 cm]{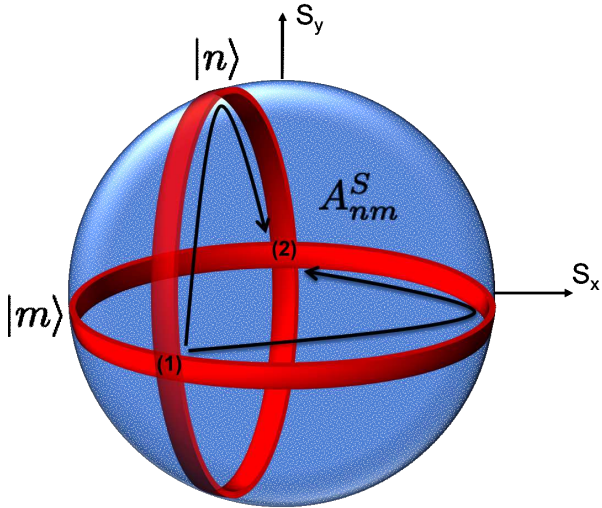}
\caption{Interference of a spin state $|n\rangle$ with respect to the quantization axis $x$ and a spin
state $|m\rangle$ with respect to the quantization axis $y$. Spin states are represented by circular bands
about their corresponding quantization axes. Their overlaps are shown by the symmetric regions $(1)$ and $(2)$,
each with area $a^S_{nm}$. Among four
possible paths from $(1)$ to $(2)$, two paths are shown with solid black curves. The surface area 
between the indicated paths on the Bloch sphere of radius $R_S$ is shown by $A^S_{nm}$. 
The transition probability between the states is determined by $P(n,m;S)=4(a^S_{nm}/2\pi R_S)\cos^2\phi^S_{nm}$, with
$\phi^S_{nm}$ being an interfence angle determined by  $A^S_{nm}$ and $R_S$.}
\label{fig:BlochSphereInterf}
\end{figure}

The spin phase-space is defined
by $(S_x,S_y,S_z)$ where the spin states can be described on the Bloch sphere of radius $R_S=\sqrt{S(S+1)}$ 
for a given total spin $S$ as shown
in Fig.~\ref{fig:BlochSphereInterf}~\cite{lassig_interference_1993}.
Each angular momentum
state can be represented by an annulus of unit width centred on a semi-classical Kramers trajectory (similarly to the WKB method) defined by 
Planck-Bohr-Sommerfeld quantization conditions set on the surface of the Bloch sphere.
Interference, mathematically encompassed into the transition probabilities $P(n,m)$, 
is geometrically understood as the intersection of two such annuli, one about the $x$-axis and the other about the $y$ one. 
This allows us to write the transition probability
in the form~\cite{lassig_interference_1993}
\begin{equation}\label{eq:dMatrixInterfForm}
\tag{S1}
P(n,m)=4\frac{a^S_{nm}}{2\pi R_S}
\cos^2{\left(\frac{A^S_{nm}}{2R_S}-\frac{\pi}{4}\right)},
\end{equation}
where $a^S_{nm}$ is the area of the intersection between the annuli
and $A^S_{nm}$ is the area bounded by the circles with $S_x=n$ and $S_y=m$ [cf. Fig.~\ref{fig:BlochSphereInterf}]. 
The general expressions of these quantities are given in Ref.~\cite{lassig_interference_1993}. 

According to Eq.~(\ref{eq:dMatrixInterfForm}), semi-classical interference in the spin 
phase-space is associated with the 
quantum interference of transition probabilities. 
Due to the additional thermal distribution weight factors in the internal energy expressions, we have a multitude of overlapping Kramers trajectories, which implies the consideration of a thermally weighted phase-space interference grid~\cite{gagen_phase-space-interference_1995}, which might result in a washing out of any oscillatory behavior (a typical signature of interference) in the work output~[cf.~Eqs.~(4) and ~(7)]. 
Moreover, care should be used to gauge the nature of the contributions to such oscillations. Indeed,  both $W_x$ and $W_{xy}$ can exhibit oscillations with integer and half-integer values of $S$, due to the
spectral difference of the unperturbed Hamiltonian in $H_\alpha^{(0)}$. However such oscillations would not be related to any quantum interference. In order to discern the quantum interference effects in the work output, we can
consider two approaches. One possibility is to measure the general overall work output and its value for $\gamma_y=0$, which would give exactly $W_x$, and infer $W_{xy}$ from the difference $W-W_x$. Alternatively, one can measure the total work output for positive and negative values of the difference $\gamma_y^H-\gamma_y^L$, labelling such values as $W_{+}$ and $W_-$ respectively. The interference term in the work output would then be inferred as $W_{xy}=(W_+-W_-)/2$.

\section{Phase space interference in the engine cycle }
\label{subsec:phaseSpaceInterference}

We have seen that work output of the engine can exhibit even-odd oscillations with respect to number of quantum workers,
which is particularly significant in the case of non-extensive working system. We can isolate the genuine
quantum interference contribution to these oscillations by using either $W_{xy}=W-W_x$ or $W_{xy}=(W_+-W_-)/2$ 
for sufficiently low number of spins where the perturbation method is satisfactory. Our objective now is to establish more
clear link between semiclassical interference picture in the spin phase space with the even-odd oscillations in the work output.
For that aim, let us consider the factors contributing to the work output expression, Eq.~(7), in the main text.

\begin{figure}[tbp]
\centering
   \subfloat[ \label{fig:DeltaPnInteg}]{\includegraphics[width=6 cm]{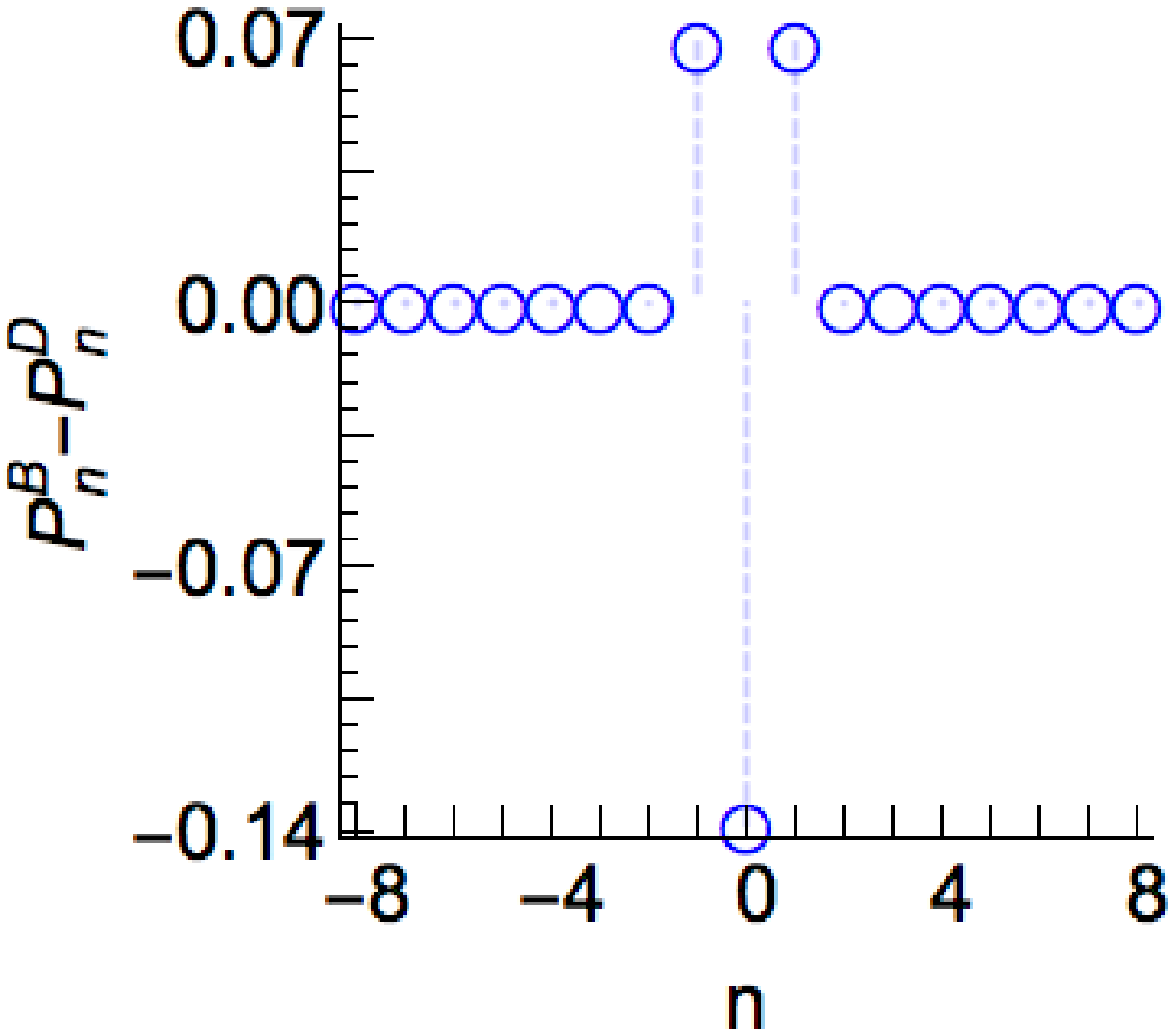}}
   \hfill
   \subfloat[ \label{fig:DeltaPnHalfInteg}]{\includegraphics[width=6 cm]{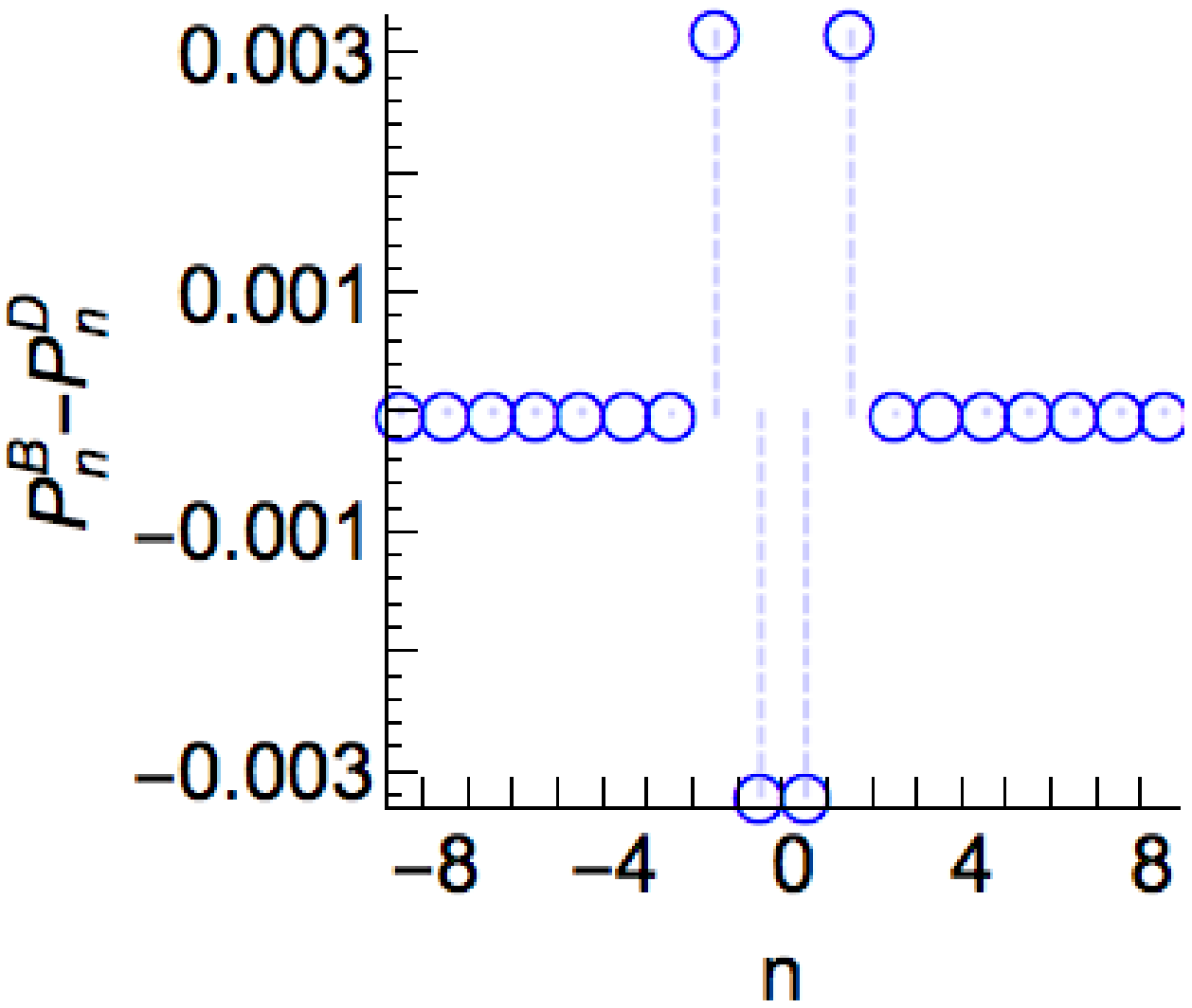}} 
\caption{Population change $P_n^B-P_n^D$ of unperturbed levels $n$ for (a) the total spin $S=8$ and (b) $S=17/2$.
Other $S$ values exhibit the similar behavior. 
We use
dimensionless parameters scaled by $\gamma_x^L$ and take the spin-spin interaction parameters as 
$\gamma_x^H=1.01$, $\gamma_x^L=1$,
$\gamma_y^H=0.01$ and $\gamma_y^L=0.02$. Temperatures of the heat baths are taken to be
$T_H=0.4$ and $T_L=0.1$. }
\label{fig:DeltaPn}
\end{figure}

The behavior of the first factor $\Delta P_n:=P_n^B-P_n^D$, which is the population change of an unterperturbed level $n$, is
calculated by the Eqs.~(5)
and plotted in Fig.~\ref{fig:DeltaPn}. Other $S$ values give
the same result that the dominant $n$ contributing to the work output are $n=0,\pm 1$ and $n=\pm 1/2,\pm 3/2$ for
integer and half-integer $S$. This is enforced by taking $T_H\ll\gamma_x^H$ and $T_L\ll \gamma_x^L$ to restrict the number
of the spin bands on the Bloch sphere forming an interference grid (cf.~\ref{fig:BlochSphereInterf}). The relations
$\Delta P_0 = 2\Delta P_{\pm 1}$ and $\Delta P_{\pm 1/2}=\Delta P_{\pm 3/2}$ holds true for other $S$, too.
\begin{figure}[h!]
	\subfloat[ \label{fig:semiclVsExactDmatHalfInteg}]{\includegraphics[width=4 cm]{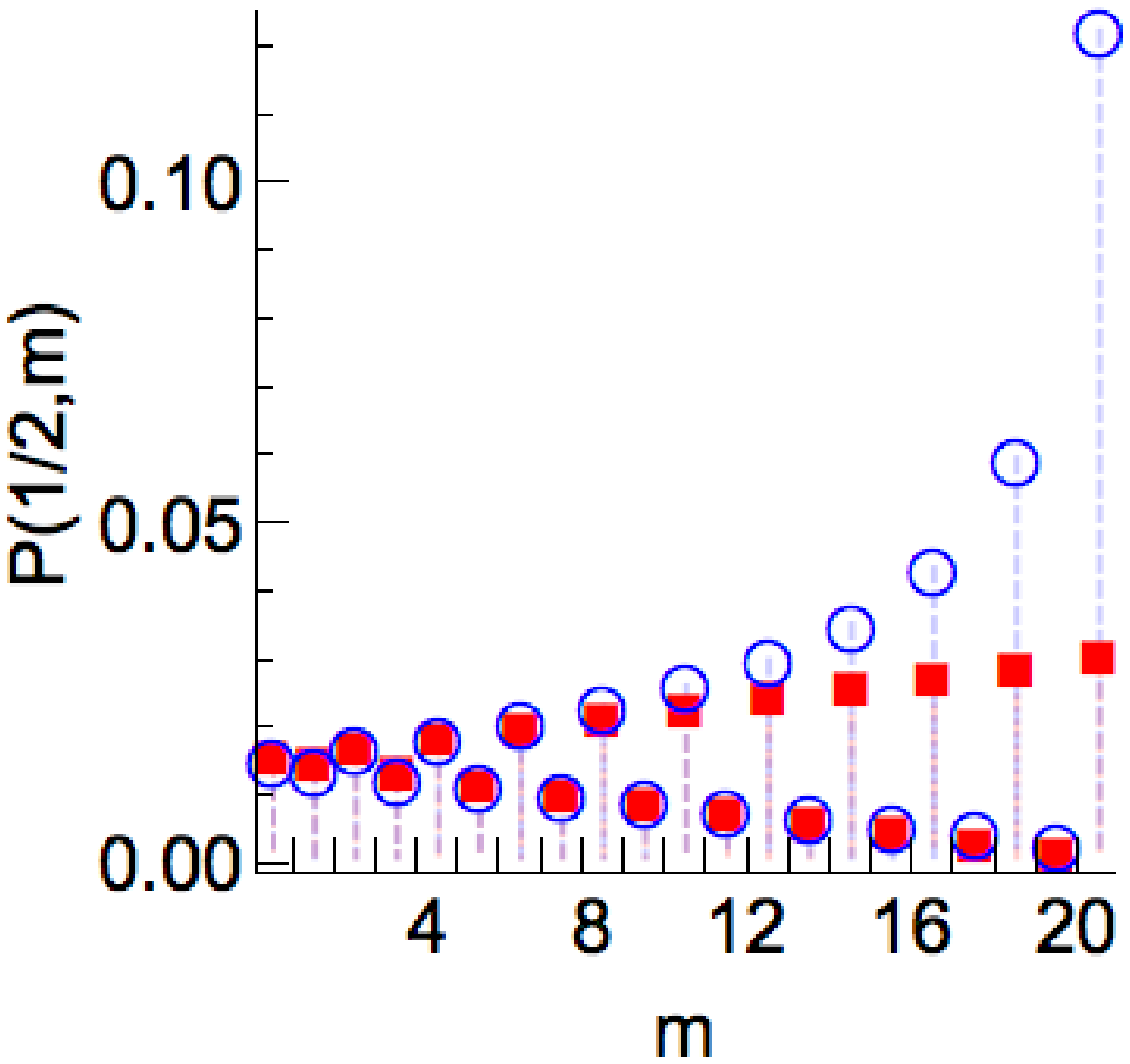}}
	\hfill
	\subfloat[ \label{fig:semiclVsExactDmatInteg}]{\includegraphics[width=4 cm]{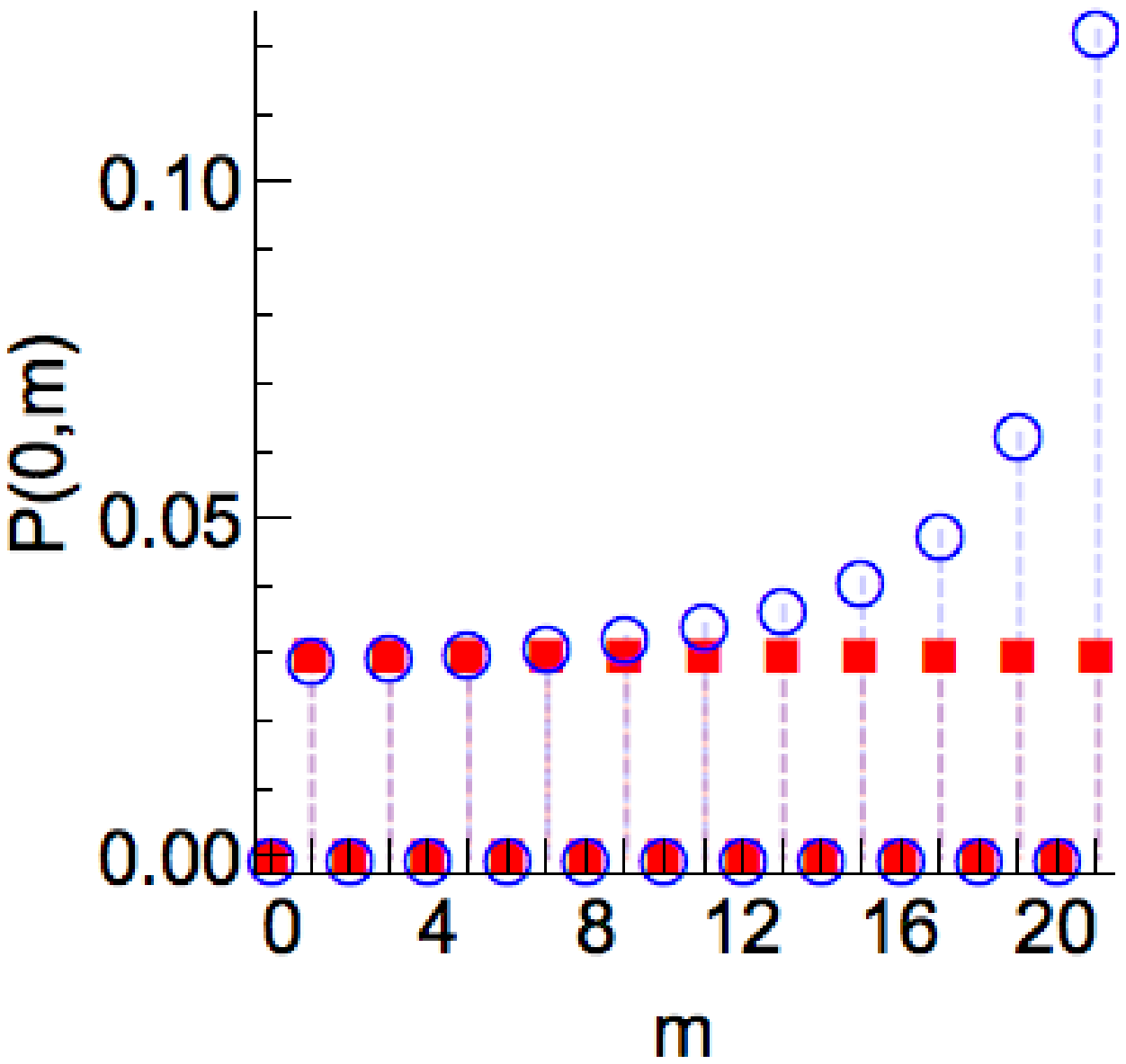}} 
	\subfloat[ \label{fig:DeltaPnmHalfInteg}]{\includegraphics[width=4 cm]{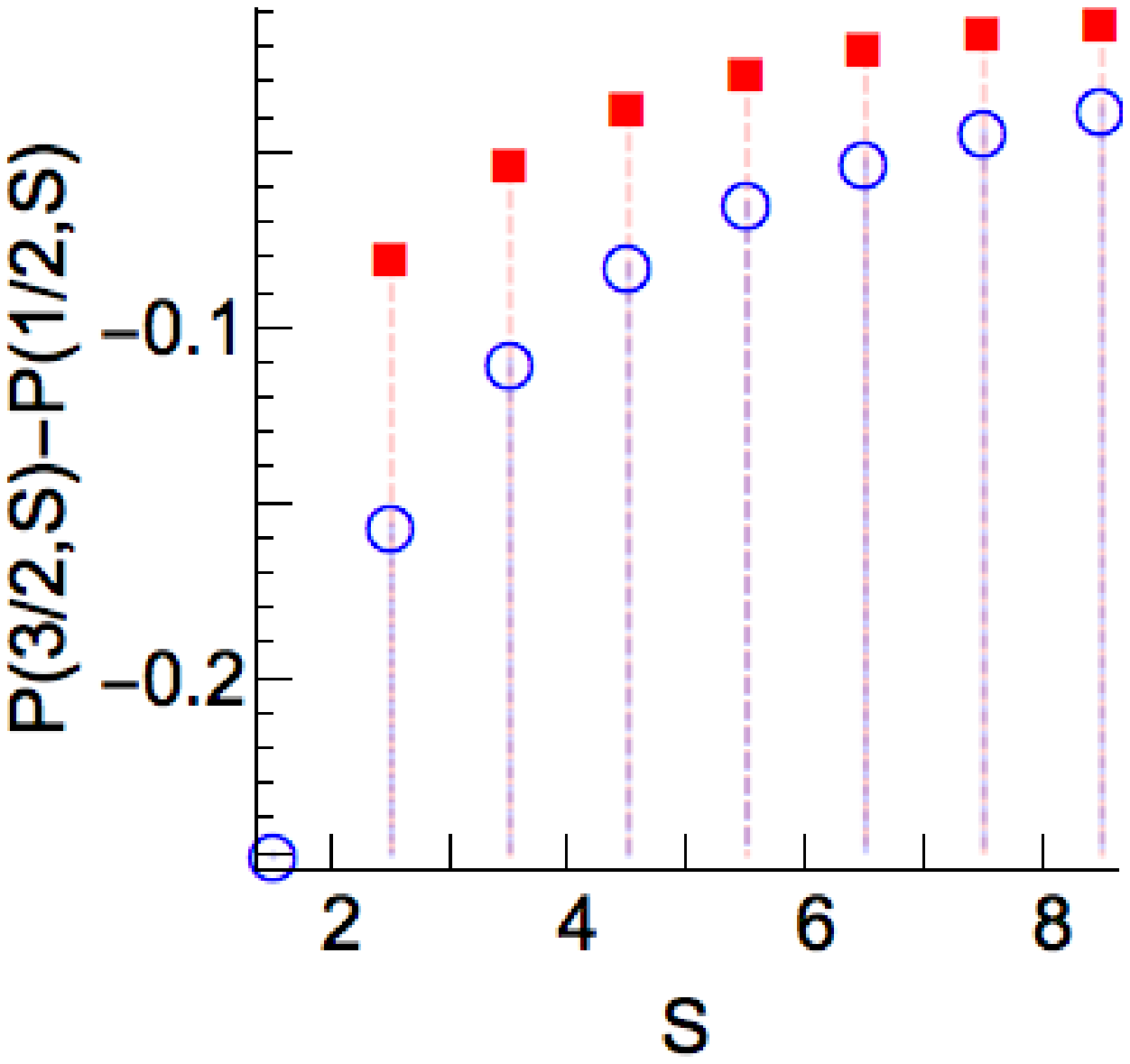}}
	\hfill
	\subfloat[ \label{fig:DeltaPnmInteg}]{\includegraphics[width=4 cm]{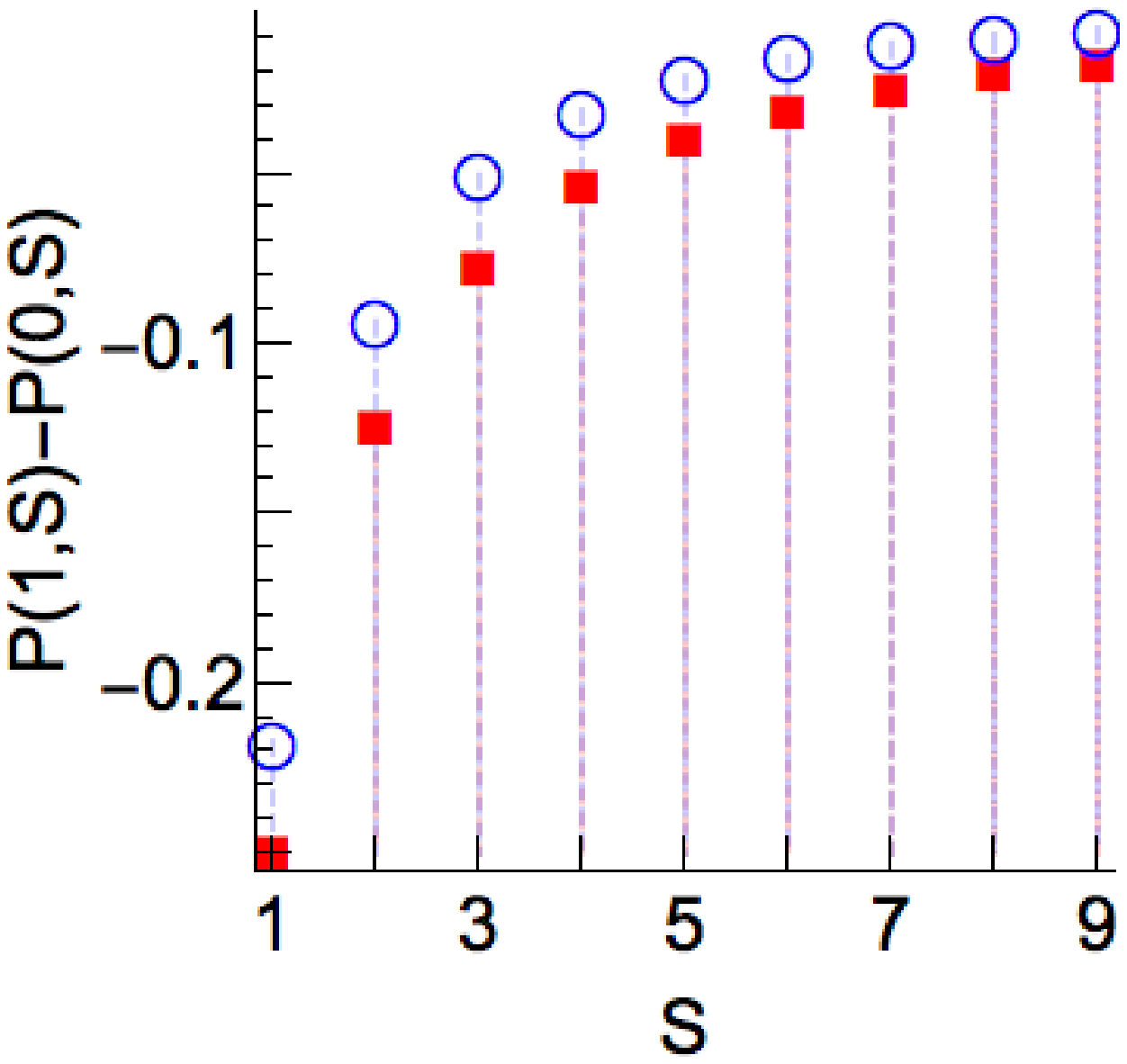}} 
	\caption{Comparision of semiclassical (red squares) and exact evaluation (blue circles) of transition probabilities $P(n,m)$ 
		between $x$-quantized spin states $|n\rangle$
		and $y$-quantized spin states $|m\rangle$ for (a) $n=0$ and total spin $S=21$ and (b) $n=1/2$ and $S=21/2$. Transition probability
		differences (c) $P(3/2,S)-P(1/2,S)$ and (d) $P(1,S)-P(0,S)$ show better agreement between the semiclassical and exact evaluations. }
	\label{fig:Pnm}
\end{figure}
The second factor is the energy change and increases quadratically with $n$. A
spectrum linear in $n$ cannot yield interference oscillations, which can be seen by the symmetry properties of populations and
transition probabilities. In contrast to $x$-quantized spin levels $|n\rangle$, which is thermally limited to low levels, 
energetically higher levels in $y$ quantized spin $|m\rangle$ participate to interference grid. Hence, we can further approximate
the dominant bands in the grid by taking $m=\pm S$. As the integer and half-integer $S$ are decoupled in the Hamiltonian, we can
deduce that the grid approximately consists of pairs of bands $(n,m)$ which are $(0,\pm S), (\pm 1,\pm S)$ and $(\pm 1/2,\pm S^\prime),
(\pm 3/2,\pm S^\prime)$, for integer $S$ and half-integer $S$, respectively. 

The transitions between the intersecting points of these bands lead to interference effects and the work output can be expressed
accordingly as
\begin{equation}
\tag{S2}
W_{xy}=4\Delta P_{k}\Delta\gamma_yS^2[(P(k,S)-P(k+1,S)],
\end{equation}
where $\Delta\gamma_y:=\gamma_y^H-\gamma_y^L$, $k=1/2$ and $k=0$ for half-integer and integer $S$, respectively.
This formula should only be used for qualitative purposes. While the consideration of $n$ by the few lower bands is a 
quite good description, the restriction of $m$ by only the highest bands $\pm S$ is a quite poor approximation. In fact $\sum_mm^2P(n,m)$ differs
from $2S^2P(n,S)$ significantly after $S>5$. However, even-odd oscillations and the essential physics are captured qualitatively
by limiting ourselves to the $m=S$ band. The other $m$ bands smoothen the oscillatory behavior, making the work
output more uniform over even and odd $S$.

\begin{figure}[tbp]
	\centering
	\subfloat[ \label{fig:Phi}]{\includegraphics[width=6 cm]{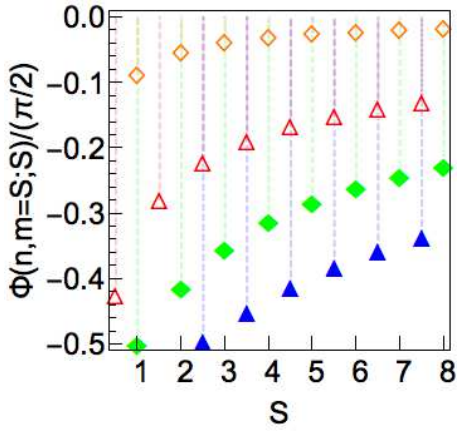}}
	\hfill
	\subfloat[ \label{fig:PhiTopView}]{\includegraphics[width=6 cm]{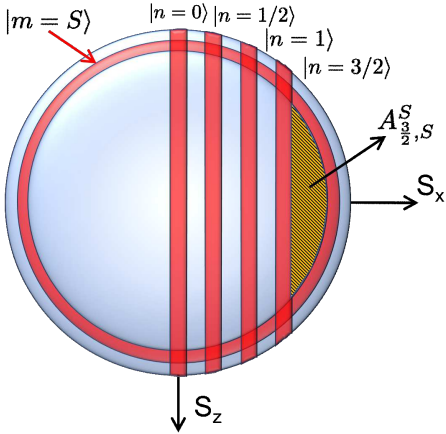}} 
	\caption{
		(a) Dependence of interference angle $\Phi(n,m=S;S)$ (scaled with $\pi/2$) on the total spin $S$. Spin state for the quantization
		axis $x$ ($y$) is denoted by $|n\rangle$ ($|m=S\rangle$).
		Plot markers of empty orange diamond, empty red triangle, filled green diamond, and
		filled blue triangle are for $n=0$, $n=1/2$, $n=1$, and $n=3/2$, respectively. (b) Top view of the Bloch sphere. Areas
		of overlap between the $|n\rangle$ and $|m=S\rangle$ decreases with $n$. The shaded area $A^S_{3/2,S}$ is the smallest.
	}
	\label{fig:PhinmS}
\end{figure}
\begin{figure}[h!]
	\centering
	\subfloat[ \label{fig:cos2}]{\includegraphics[width=6 cm]{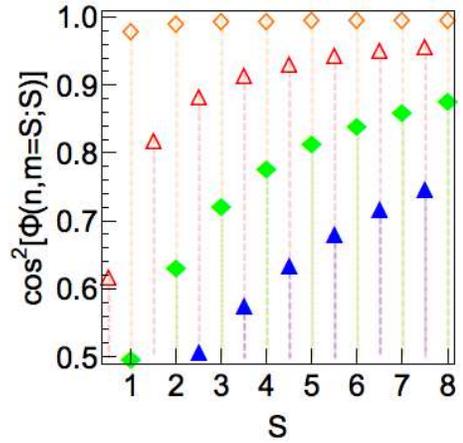}}
	\hfill
	\subfloat[ \label{fig:pnmfull}]{\includegraphics[width=6 cm]{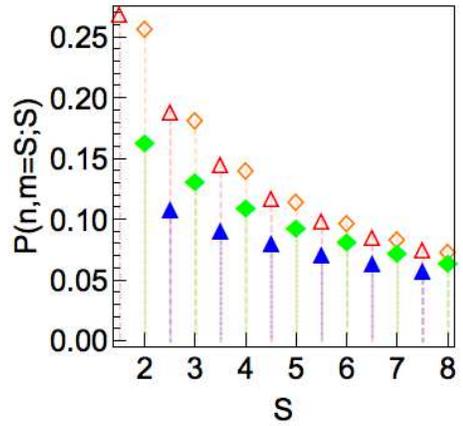}} 
	\caption{
		Dependence of (a)  interference cosine (squared) $\cos^2{\Phi(n,m=S;S)}$ and (b) $P(n,m=S;S)$ on the total spin $S$. 
		Plot markers of empty orange diamond, empty red triangle, filled green diamond, and
		filled blue triangle are for $n=0$, $n=1/2$, $n=1$, and $n=3/2$, respectively.
	}
	\label{fig:delPexplained}
\end{figure}

We use the semiclassical formula in Eq.~(\ref{eq:dMatrixInterfForm}) for $P(n,m)$ to calculate the work output. The difference between the semiclassical
evaluation and the exact value increases towards larger $m$, both for the cases of half-integer (cf.~Fig.~\ref{fig:semiclVsExactDmatHalfInteg})
and integer total spin (cf.~Fig.~\ref{fig:semiclVsExactDmatInteg}). 
The particular qualitative behaviors we look for in the transition probability differences are
plotted in Figs.~\ref{fig:DeltaPnmHalfInteg} and~\ref{fig:DeltaPnmInteg}.

The negativity of the population differences can be associated with the behavior of the interference angle in Eq.~(\ref{eq:dMatrixInterfForm}) 
which is $\Phi(n,m;S):=A^S_{nm}/2R_S-\pi/4$. The largest overlap area is between the states $|n=0\rangle$ and $|m=S\rangle$.
It is analytically known that associated $\Phi(0,S;S)=0$~\cite{lassig_interference_1993}. Indeed, the surface area of a spherical cap
on a sphere of radius $R$ is given by $\pi Rh$, with $h$ is the height of the cap. We find $A^S_{0S}=\pi R_S(R_S-S)$. For $S\gg 1$
we have $R_S\sim S+1/2$ so that $A_S\sim \pi R_S/2$ which gives $\Phi(0,S;S)\sim 0$. Other areas between the corresponding bands
gets smaller and $\Phi\rightarrow -\pi/4$. These deductions are verified in Figs.~\ref{fig:Phi} and~\ref{fig:PhiTopView}.

We plot $\cos^2\Phi(n,S;S)$ with respect to $S$ in Fig.~\ref{fig:cos2}. The differences between the interference cosines 
for $n=0$ and $n=1/2$ is smaller than those for $n=1$ and $n=3/2$. The amplitude of the cosines, $\sim 2/\pi R_S$ regularly
decrease with $S$ and do not change the hierarchy $P(0,S;S)\lesssim  P(1/2,S^\prime;S^\prime)<P(1,S;S)<P(3/2,S^\prime;S^\prime)$ 
we would deduce by the examination of the interference angles and the cosines, as verified by 
Fig.~\ref{fig:pnmfull}. Here $S$ and $S^\prime$
are integer and half-integer total spin vlaues neighboring each other on the real axis. Hence,
we conclude that there are oscillations in the transition probability differences contributing to the work output with respect to
integer and half-integer $S$ (cf.~Figs.~\ref{fig:DeltaPnmHalfInteg} and~\ref{fig:DeltaPnmInteg}). This however would suggest a larger
interference contribution from odd number of spins as $P(1/2,S^\prime;S^\prime)-P(3/2,S^\prime;S^\prime)>P(0,S;S)-P(1,S;S)$. According
to Fig.~\ref{fig:DeltaPn} however, thermal populations in lower levels have the relation $|\Delta P_0|\gg|\Delta P_{1/2}|$. This thermal boost
enhances the interference contribution of even number of spins to the work output more than those of the odd number of spins. Finally,
having seen that $P(k,S)-P(k+1,S)$ is positive, the 
negativity of the $\Delta P_k$ can be compansated by taking the $\Delta\gamma_y<0$ so that the interference is tuned to constructive
contribution to work output.

%

\end{document}